\documentclass[11pt,twoside]{article}
\usepackage{graphicx}
\usepackage{amsmath}
\usepackage{asp2010}

\resetcounters

\begin{document}

\title{Oscillations and Surface Rotation of Red Giant Stars}
\author{Christina Hedges$^{1,2}$, Savita Mathur$^{1,3}$, Michael J. Thompson$^1$, and Keith B. MacGregor$^1$
\affil{$^1$High Altitude Observatory, NCAR, P.O. Box 3000, Boulder, CO 80307, USA}
\affil{$^2$School of Physics and Astronomy, University of Birmingham, Edgbaston, Birmingham, B15 2TT, UK}
\affil{$^3$Laboratoire AIM, CEA/DSM-CNRS-Universit\'e Paris Diderot; IRFU/SAp, Centre de Saclay, 91191 Gif-sur-Yvette Cedex, France}}

\begin{abstract}

More than 15000 red giants observed by Kepler for a duration of almost one year became public at the beginning of this year. We analysed a subsample of 416 stars to determine the global properties of acoustic modes (mean large separation and frequency of maximum power). Using the effective temperature from the Kepler Input Catalog, we derived a first estimation of the masses and radii of these stars. Finally, we applied  wavelets to look for signature of surface rotation, which relies on the presence of spots or other surface features crossing the stellar visible disk. 
\end{abstract}

\section{Introduction}
The last decade has seen an exponential increase in the interest and knowledge of red giants. The structure of red giant stars is less well known than that of solar-like stars. Thanks to the Kepler and CoRoT missions seismologists are able to do more statistical analysis on these evolved stars \citep{2011A&A...525A.131H} and to unveil little by little their core structure \citep[e.g.][]{2011Natur.471..608B} and their internal dynamics \citep{2012Natur.481...55B,2012A&A...548A..10M}. The study of their regular modes and mixed modes, which arise from the coupling between the cavities of acoustic and gravity modes, leads to insights into the internal structure of these stars.

Further, the study of surface rotation by wavelet analysis can show how active the surface of these stars can be. Indeed, recently spectropolarimetric observations have shown hints of magnetic field in giants \citep[e.g.][]{2012A&A...541A..44K}. In this work, we analyse data for several hundreds of red giants observed by Kepler.

Through the combination of asteroseismic analysis probing the interiors of stars and the wavelet analysis of the surface features we can better piece together a picture of the structure of evolved stars, furthering our understanding. This can give us insights not only into the physical structure of the stars but also show us what levels of activity may or may not be preserved in red giant branch stars.

\section{Data}

Kepler has been monitoring more than 100,000 stars for several years, producing high quality light curves. Some of these stars are known to be red giant stars. These stars have been picked out for further study in order to see how they differ from less evolved stars. We chose to analyse 416 random public red-giant stars that are more or less evolved. These stars had data series' that spanned approximately two years of observation. The data were corrected for outliers, jumps, and trends and then concatenated following \cite{2011MNRAS.414L...6G}. These data were used to produce Power Density Spectra as shown in Figure 1 in order to look for their asteroseismic properties.

\begin{figure}[htbp]
\begin{center}
\includegraphics[width=8cm]{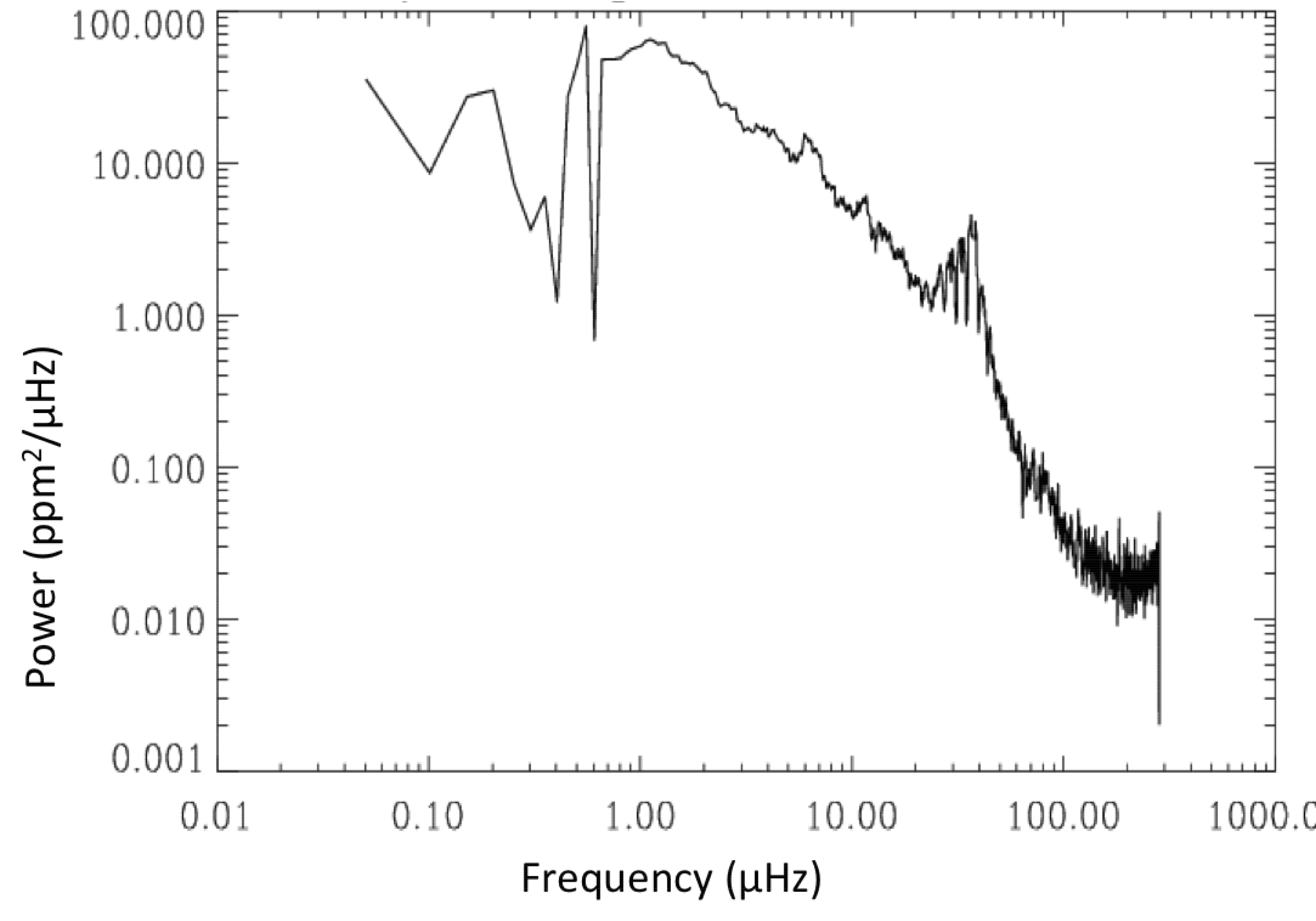}
\caption{Example of a Power Density Spectrum of a light curve from a red giant observed by Kepler. The envelope of modes can be seen at approximately 50\,$\mu$Hz.}
\label{fig1}
\end{center}
\end{figure}

\section{Global acoustic parameters}

\subsection{Developing the pipeline}

We developed a pipeline to process the Kepler data automatically and output useful plots and data files for further analysis. The processes are as follows, starting with a raw light curve from Kepler in time and flux:
\begin{enumerate}
\item Convert to appropriate units and discard infinite points
\item Generate a Power Density Spectrum (PDS) using a Lomb-Scargle Periodogram after applying a 1 day filter
\item Smooth the PDS
\item Find the region with the most power, (i.e. the envelope of modes) excluding the low frequency regime
\item Compute the Power Spectrum of the Power Spectrum \citep[e.g.][]{2010A&A...511A..46M}
\item Measure values for the large frequency spacing and the frequency of maximum power  
\item Check the stars that do not follow the general trend of our sample and re-do the analysis
\end{enumerate}

By combining the large frequency spacing and the frequency of maximum power with the effective temperature from the Kepler Input Catalogue \cite[KIC,][]{2011AJ....142..112B} we obtain a first estimate of mass and radius using scaling relations \citep{kjeldsen95}.

\subsection{Results}

\begin{figure}[htbp]
\begin{center}
\includegraphics[width=8cm]{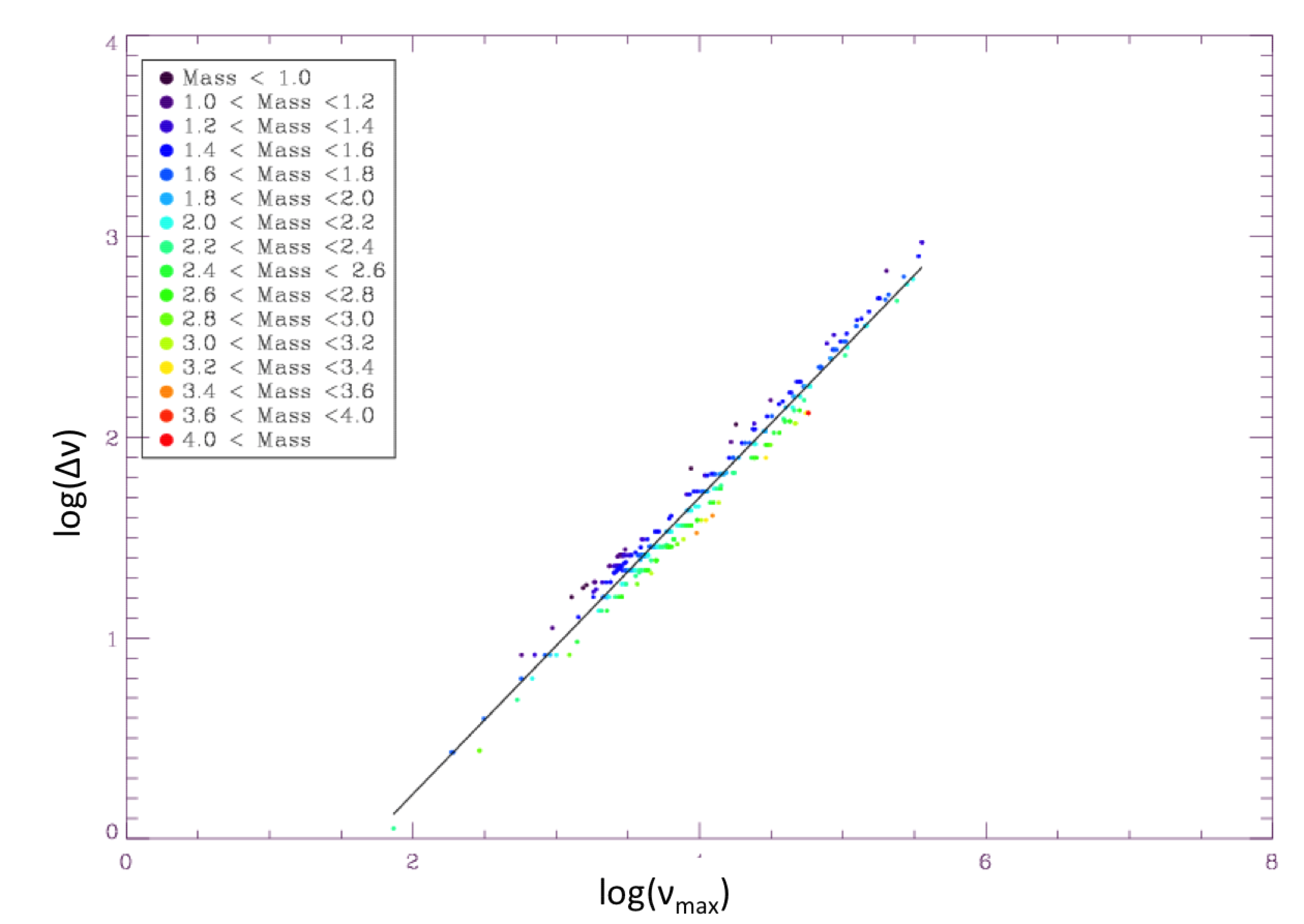}
\caption{Mean large spacing against frequency of maximum power obtained by our pipeline. The colour code represents the  mass (in terms of solar mass) of the stars derived with the scaling laws. The black line is the linear fit of the data. This was used to asses outlier points to be ignored for further analysis.}
\label{fig2}
\end{center}
\end{figure}

Having developed this pipeline, we applied it to our sample of 416 red giants. The pipeline was over 80\% effective at extracting the global parameters of the acoustic mode of the sample of red giant data studied. Figure~\ref{fig2} shows the large frequency spacing against the frequency of maximum power. We clearly see a linear trend between these two quantities in the log-log scale. By fitting this trend we obtain the following relationship that agrees well with \cite{2010ApJ...713L.176B} and \cite{2011ApJ...743..143H}:

\begin{equation}
\Delta \nu = 0.246 \nu_{\rm max}^{0.739}.
\end{equation}

\noindent This relationship is dependent on the mass of the star as can be noticed in Figure~\ref{fig2}. This dependence found here was consistent with other papers from the field \citep{2010A&A...517A..22M}. The plot also shows an agglomeration of points at $\nu_{\rm max}$ around 30-40$\mu$Hz. This corresponds to the ``Red Clump'' where red giants have started burning helium in their cores, marking another step in their evolution.

\begin{figure}[htbp]
\begin{center}
\includegraphics[width=8cm]{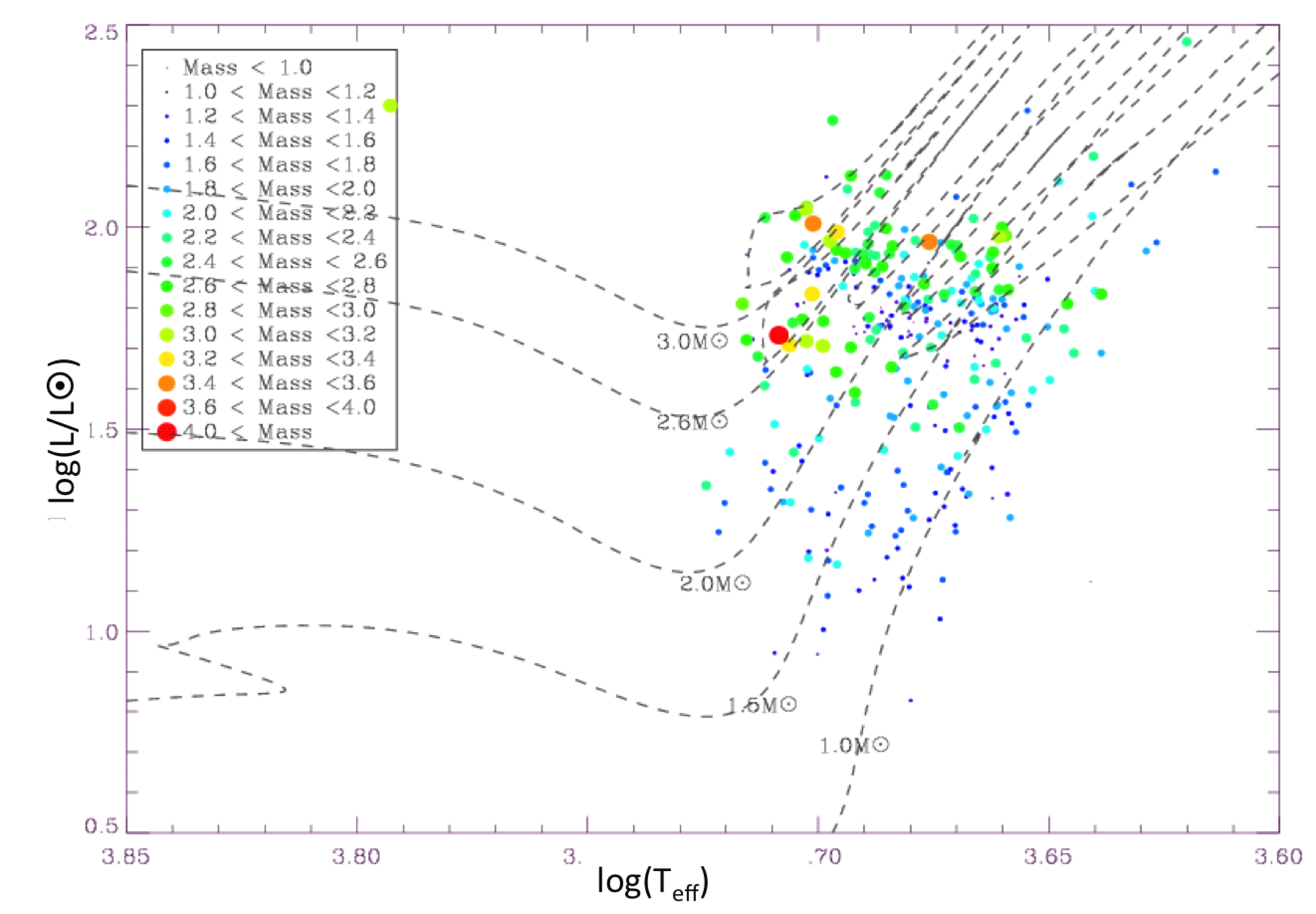}
\caption{HR diagram (Luminosity vs Teff) including evolutionary tracks based on metallicity. The magnitudes of the masses are slightly skewed as discussed above. }
\label{fig3}
\end{center}
\end{figure}

\noindent We studied how the observations match with models by representing our pipeline results in an HR diagram (see Figure~\ref{fig3}). Due to the wide variation in crucial factors from observations of these stars the temperatures given by the KIC have errors of 200K. This produces  large uncertainties in the values for mass and radius. This led to an amount of skewing in the mass parameter leading to overly high masses. However the general trend shows what would be expected in red giant evolution.

\section{Study of the surface rotation}

\begin{figure}[htbp]
\begin{center}
\includegraphics[width=8cm]{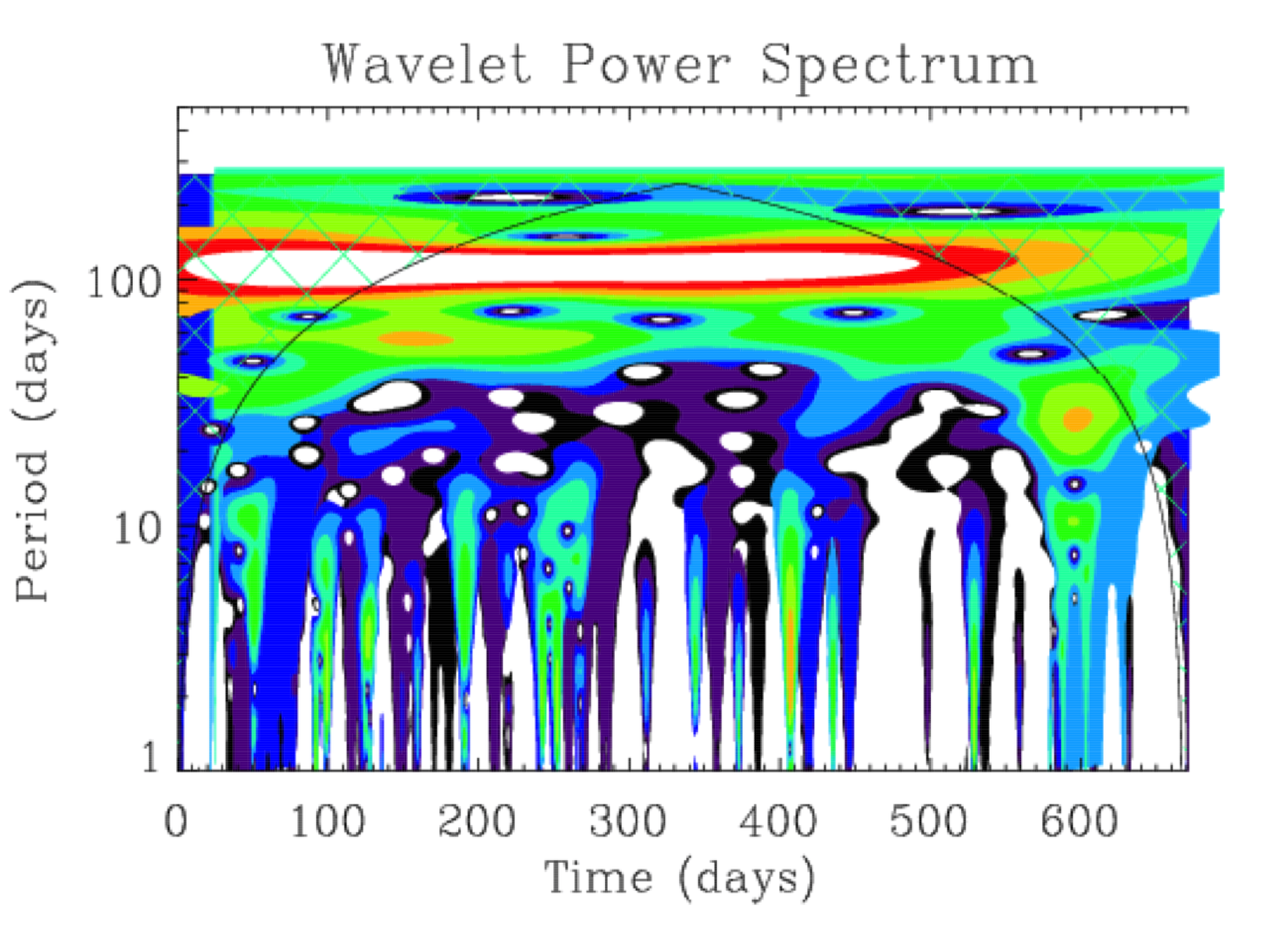}
\caption{Example of a Wavelet Power Spectra with surface rotation being shown. The hatched lines indicate the "cone of influence" where period is too long for the wave to have repeated four times. Therefore results in the hatched area are disregarded. The red contour at 112 days shows the rotation. The colouring in the graph shows blue and green as low power and orange, red and white as high power. It is normal to find there is some high power of fit when using short period wavelets.}
\label{fig4}
\end{center}
\end{figure}

The method of Wavelet analysis \citep{1998BAMS...79...61T} can be used to perform a time-frequency analysis to find periodicity as a function of time. In terms of the time series obtained by Kepler a contour plot can be produced to show if there is any period that fits an oscillation of any part of the light curve. This can tell us whether the star is rotating or not as long as there is some bright or dark feature on the surface moving in and out of view.
Wavelet analysis is where a mother wavelet, usually a periodic function that is modified by a function like a Gaussian, (e.g. a Morlet function,) is iteratively fit to the time series. This function has its period changed and iterates to find the best fit. The correlation between the mother wavelet and the time series gives the Wavelet Power Spectrum. The highest power is obtained where the period of the mother wavelet matches the observations. An example of a WPS is shown as a contour plot in Figure~\ref{fig4}.

In order for some periodicity to be measured there must be some identifiable change on the surface as it rotates. There must be some dark or bright part of the star that is constantly turning into and out of view. If the star is perfectly uniform no change will be detected and it will not appear to rotate.
Rotation can not always be detected. Several factors affect the result whether the star appears to be rotating or not: inclination, inactivity, a very long rotational period (Kepler light curves are approximately 2 years long and a detection requires at least 4 rotations have been observed) and the latitude of feature amongst other things. When all these problems are taken into account only a few stars will show useful rotation measurements.
The Kepler satellite itself has certain properties that are periodical and can manifest in the Wavelet analysis. These can be related to the downlink of the Kepler data every month or the rollover of the satellite every 3 months so that the solar panels keep facing the Sun. These lead to an artefact at 90 and 170 days. We disregarded the stars where these periods were obtained.  We visually inspected the data and 14 stars showed clear rotation profiles. These stars all occurred within the same series of KIC number, suggesting there is something linking these stars. It was investigated whether these stars shared the same module number but they were found to be different.

Where a contour was present that indicated rotation the Wavelet Power Spectra are indicative of some feature on the surface of the star. When looking into the detail of the periodicity we find these high power regions correspond to a negative change in brightness, leading to the belief there is a dark patch on the surface of the star.

These dark patches may be indicative of some activity in the red giant stars such as ``Starspots''. This is interesting as red giants are thought to be largely inactive due to their highly evolved state. In some of the contour plots that have successfully shown rotation the contour showing periodicity has been ``broken'' and was present for a few hundred days. This indicates the feature is not permanent but is lasting much longer than an average sunspot.
Another feature of the contour plots is the width  of the contour showing rotation. The period spans more than 20 days in range. There are clearly thinner contours around the same area showing the rotation contour is much wider than the period resolution of the plot. This implies that this is not simply a widening of the contour due to a resolution limit but that the feature itself is apparent across many different values of period. This shows that parts of the star must be rotating with different periods, i.e. there is latitudinal differential rotation on the surface of this star where this feature is present. Similar plots for the Sun can be found in Hempelman, 2003, A$\&$A, 399, 717 which shows sets of wide contours that vary in intensity over the course of a solar cycle.

\section{Conclusions}

With our pipeline, we analyzed a large number of red giants to measure the global parameters of acoustic modes. We derived the relation between the mean large separation and the frequency of maximum power.
Mass and radius can be  estimated from the determination of acoustic-mode global parameters.
Using wavelet analysis an approximation for the period of the surface rotation can be found. The rotation was of the order of 100 days in the cases found here. 
The Wavelet Power Spectra show that the surface rotation is not always present over the 2 year period, suggesting that features may be present only some of the time. 
The features in the Wavelet Power Spectra tend to be strong with a wide range of periods, supporting the theory that red giants rotate differentially. 
While fine structure cannot be resolved in the contour plots currently the presence of dark features in them appears to support the theory that stars do continue to have some activity after turning off the main sequence in some cases.

\section{Acknowledgements} C. H. acknowledges the support of the REU program from NCAR and the NSF. NCAR is partially funded by the National Science Foundation. SM acknowledges the support of the European Community's Seventh Framework Programme (FP7/2007-2013) under grant agreement no. 269194 (IRSES/ASK), the CNES, and Tokyo University.

\bibliographystyle{asp2010} 
\bibliography{/Users/Savita/Documents/BIBLIO_sav.bib}

\end{document}